\documentclass[pre,twocolumn,showpacs]{revtex4}

\usepackage{graphicx}
\usepackage{bm}
\usepackage{amssymb}

\begin{document}
\title{Fractional dynamics from the ordinary Langevin equation}
\author{A. A. Stanislavsky}
\email{alexstan@ri.kharkov.ua} \affiliation{Institute of Radio
Astronomy, 4 Chervonopraporna~St., Kharkov 61002, Ukraine}

\date{\today}

\begin{abstract}
We consider the usual Langevin equation depending on an internal
time. This parameter is substituted by a first passage time of a
self-similar Markov process. Then the Gaussian process is parent,
and the hitting time process is directing. The probability to find
the resulting process at the real time is defined by the integral
relationship between the probability densities of the parent and
directing processes. The corresponding master equation becomes the
fractional Fokker-Planck equation. We show that the resulting
process has non-Markovian properties, all its moments are finite,
the fluctuation-dissipation relation and the H-theorem hold.
\end{abstract}
 \maketitle

\section{Introduction}
The Langevin equation is a powerful tool for the study of
dynamical properties of many interesting systems in physics,
chemistry and engineering \cite{1,2}. The success of the approach
rests on the description of macroscopic quantities starting from
microscopic dynamics, where the effect of fast degrees of freedom
(heat bath in statistical physics and solid state physics, short
wavelength modes in meteorological and climate models, etc) can be
often taken into account by noise \cite{3,4,5,6}. Gaussian noise
leads to normal diffusion with a mean square displacement that
grows linearly in time and to an exponential relaxation.
Equivalently, the phenomena can be also described by the ordinary
Fokker-Planck equation (FPE) for the time evolution of the
probability density of the random processes.

However, many systems exhibit anomalous behavior in their
transport and relaxation properties \cite{7,8}. Anomalous
diffusion has the mean square displacement increasing as a
(non-linear) power law in time, and anomalous relaxation shows a
slow power law decay in the long-time limit. The attempt to state
a dynamical foundation in statistical physics, as well as the
great interest in understanding the physical mechanism leading to
anomalous diffusion/relaxation, calls into being the generalized
Langevin equations \cite{9,10}. The generalizations affect either
the equation form itself (for example, via the memory kernel)
or/and the character of correlations in the fluctuating force
\cite{11,12,13}. The way to the description of anomalous
diffusion/relaxation is not unique. In this paper we show that the
ordinary Langevin equation can result in anomalous
diffusion/relaxation owing to the fact that the temporal degree of
freedom becomes stochastic. The approach clarifies a microscopic
derivation and interpretation for the fractional FPE. The ordinary
Langevin equation is a particular case of the new model.

The paper is organized as follows. In Sec. \ref{par2} we introduce
the concept of the stochastic time clock. The new clock (random
process) generalizes the deterministic time clock of the ordinary
Langevin equation and governs by the random process described by
the stochastic differential equation. The directing process arises
from a self-similar $\alpha$-stable random process of temporal
steps. Using properties of the stochastic time clock, in Sec.
\ref{par3} we write the corresponding master equation with the
fractional derivative of time. To know the solution of the usual
FPE with a time-independent kernel, one can find immediately the
solution of its fractional generalization via the integral
relation. The ordinary Langevin equation has a stationary state.
The same feature remain valid, when passing to the stochastic time
clock. For this case, Sec. \ref{par4} is devoted to the
fluctuation-dissipation relation and the H-theorem. The
randomization of time clock can be also applicable for the general
kinetic equation (Sec. \ref{par5}). The fact is illustrated on a
concrete example, the relaxation in a two-state system. We end the
paper with a short summary in Sec. \ref{par6}.

\section{Stochastic time arrow}\label{par2}
The main feature of time is its direction. Time is only running
from the past to the future. In our consideration we intend to
save the property of time. For the ordinary Langevin equation the
time variable is deterministic. Now set this variable as an
internal parameter $\tau$. The motion of a point particle of
velocity ${\bm V}(\tau)$ in a thermal bath is determined by a
viscous friction $\gamma{\bm V}$ and random collisions ${\bm
W}(\tau)$, by means of
\begin{equation}
d{\bm V}(\tau)=-\gamma{\bm V}(\tau)\,d\tau+d{\bm W}(\tau).
\label{eq1}
\end{equation}
As usual, ${\bm W}(\tau)$ is a Wiener process with zero mean and
variance per unit of time equal to $2D$. Let us randomize the time
clock of the process ${\bm V}(\tau)$. No every random process is
suitable for our goal. First of all the appropriate process must
be strictly non-decreasing. Assume that the time variable is a sum
of random temporal intervals $T_i$. Let $T_i$ be independent
identically distributed variables. It is not necessary to know the
exact form of their probability distribution. Their belonging to
the strict domain of attraction of a $\alpha$-stable distribution
($0<\alpha<1$) is quite enough. The parameter restriction
$0<\alpha<1$ arises from the need to keep the random time steps
$T_i$ as non-negative random variables. The sum of random
variables $n^{-1/\alpha}(T_1+\cdots+T_n), n\in\mathbb{N}$
converges in distribution to the $\alpha$-stable one. As has been
shown in \cite{14}, there exists the limit of the following
process $T^{\Delta\tau}(\tau)=\{\lfloor\tau/\Delta\tau\rfloor+
1\}^{-1/\alpha}\sum_{i=1}^{\lfloor\tau/\Delta\tau\rfloor+1}T_i$
under $\Delta\tau\to 0$, where $\tau$ is the internal time
separated on discrete values with a step $\Delta\tau$, and
$\lfloor a \rfloor$ denotes the integer part of $a$. Let us now
make use of the limit passage from ``discrete steps'' to
''continuous ones''. The new process satisfies the relation
$T(\tau)$ ${d}\atop =$ $\tau^{1/\alpha}T(1)$, where ${d}\atop =$
means equal in distribution. The position vector of a walking
particle at the true time $t$ is defined by the number of jumps up
to time $t$. This discrete counting process is
$N_t=\max\{n\in\mathbb{N}\mid \sum_{i=1}^nT_i\leq t\}$. The
continuous limit of the discrete counting process $\{N_t\}_{t\geq
0}$ is the hitting time process $S(t)=\inf\{x\mid T(x)>t\}$
\cite{14}. The hitting time $S(t)$ is called also a first passage
time. For a fixed time it represents the first passage of the
stochastic time evolution above this time level. The random
process $S(t)$ is just non-decreasing and depends on the true time
$t$. We choose it for a new time clock (stochastic time arrow),
assuming its statistical independence on the random variable $\bm
V$.

\begin{figure}
\includegraphics[width=8.6 cm]{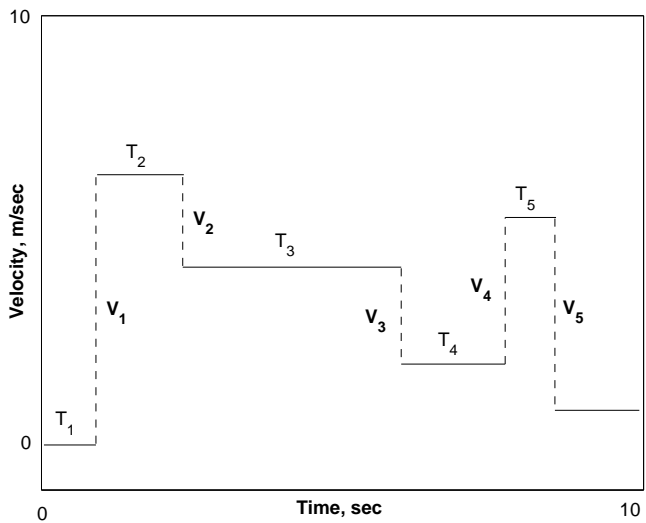}
\caption{\label{fig1}Single realization of a continuous time
random walk with random waiting-times $T_i$ between successive
random jumps of velocity $V_i$.}
\end{figure}

Although the random process $S(t)$ is self-similar, it has neither
stationary nor independent increments, and all its moments are
finite \cite{14,15}. This process is non-Markovian, but it is
inverse to the continuous limit of a Markov random process of
temporal steps $T(\tau)$, i.\ e.\ $S(T(\tau))=\tau$. The
analytical form of the probability density of the random variable
$S(t)$ can be calculated as follows. According to \cite{14}, the
expectation $\langle e^{-vS(t)}\rangle=\int^\infty_0
dx\,e^{-vx}\,p^S(t,x)$ is equal to the Mittag-Leffler function
$E_\alpha (-vt^\alpha)$. After the Laplace transform of the
Mittag-Leffler function with respect to $t$, the expectation can
be easily inverted analytically with respect $v$. Then the
probability density of the process $S(t)$ is written as
\begin{equation}
p^{S}(t,\tau)=\frac{1}{2\pi i}\int_{Br} e^{ut-\tau
u^\alpha}\,u^{\alpha-1}\,du\,, \label{eq2}
\end{equation}
where $Br$ denotes the Bromwich path. This probability density
characterizes the probability to be at the internal time $\tau$ on
the real time $t$. After the variable transform $ut\to u$ and
denoting $z=\tau/t^\alpha$, the function $p^{S}(t,\tau)$ takes the
form $t^{-\alpha}F_\alpha(z)$, where $F_\alpha(z)=\frac{1}{2\pi
i}\int_{Br}e^{u-zu^\alpha}\,u^{\alpha-1}\,du$. On deforming the
Bromwich path into the Hankel path, we find the Taylor series of
the function $F_\alpha(z)$, i.\ e.\
\begin{equation}
F_\alpha(z)=\sum_{k=0}^\infty\frac{(-z)^k}{k!
\Gamma(1-\alpha-k\alpha)}\,,\label{eq2a}
\end{equation}
where $\Gamma(x)$ is the usual Gamma function. Since the radius of
convergence of the power series (\ref{eq2a}) can be proven to be
infinite for $0<\alpha<1$, the function $F_\alpha(z)$ is entire in
$z$. Thus, the exchange between the series and the integral in the
calculations of the Taylor series (\ref{eq2a}) is quite
legitimate. The Laplace image of the function $F_\alpha(z)$ is
expressed in terms of the Mittag-Leffler function
\begin{displaymath}
\int^\infty_0e^{-z\zeta}\,F_\alpha(z)\,dz=E_\alpha(-\zeta)\,,\quad
z>0.
\end{displaymath}
Feller conjectured and Pollard proved in 1948 that the
Mittag-Leffler function $E_\alpha(-x)=\sum_{n=0}^\infty
(-x)^n/\Gamma(1+n\alpha)$ is completely monotonic for $x\geq 0$,
if $0<\alpha\leq1$. Moreover, $E_\alpha(-x)$ is an entire function
of order $1/\alpha$ for $\alpha>0$ \cite{15a}. Hence, by Feller
\cite{15a}, one can conclude that the function $F_\alpha(z)$ is
non-negative in $z>0$. Taking into account the normalization
relation $\int_0^\infty F_\alpha(z)\,dz=1$, the function
$p^{S}(t,\tau)$ is just a probability density. If $\alpha=1/2$,
from the series (\ref{eq2a}) it is easy to recognize the
well-known Gaussian function $F_\alpha(z)=\pi^{-1/2}\exp(-z^2/4)$.
It should be observed here that the basic Cauchy and Signaling
problems of the time fractional diffusion-wave equation can be
expressed in terms of the function $F_\alpha(z)$ (see more details
in \cite{15b}).

\section{Fractional Fokker-Planck equation}\label{par3}
The Langevin equation directed by the stochastic time clock $S(t)$
can be written in the form
\begin{equation}
d{\bm V}(S(t))=-\gamma{\bm V}(S(t))\,dS(t)+d{\bm W}(S(t)).
\label{eq3}
\end{equation}
The statement is quite justified. The process ${\bm V}(S(t))$ is a
continuous martingale, and the directing process $S(t)$ is a
continuous submartingale with respect to an appropriate filtration
\cite{15}. All the moments of both parent and directing processes
are finite. The same concerns the process ${\bm V}(S(t))$. Now the
random walk of a particle is defined by two Markov processes,
random waiting times $T(\tau)$ between random jumps ${\bm
V}(\tau)$. The discrete example of such a walk is shown in
Fig.~\ref{fig1}. A similar approach, i.\ e.\ the modeling of
anomalous diffusion by two independent random processes indexed by
a common continuous parameter, has been already suggested in
\cite{16}. However, the inverse process to the time evolution was
not completely defined. It was not recognized as a first passage
process. No direct relationship between Eqs. (\ref{eq1}) and
(\ref{eq3}) was found. This is also a main difference between the
subject of \cite{17} and our paper.

The resulting process ${\bm V}(S(t))$ is subordinated to ${\bm
V}(t)$, called the parent process, and is directed by $S(t)$,
called the directing process \cite{15a}. The hitting time process
$S(t)$ just satisfies necessary properties imposed on any
directing process (independent, positive and non-decreasing). The
directing process is often referred to as the randomized time or
operational time. In other words, the subordinated process ${\bm
V}(S(t))$ is obtained by randomizing the time clock of a random
process ${\bm V}(t)$ using a random process $S(t)$. In compliance
with \cite{15a}, the probability density of the random process
${\bm v}_t={\bm V}(S(t))$ is expressed by the integral relation
\begin{equation}
p^{\,v_t}({\bm V}, t)= \int^\infty_0p^V({\bm
V},\tau)\,p^S(t,\tau)\,d\tau\,, \label{eq4}
\end{equation}
where $p^V({\bm V}, \tau)$ represents the probability to find the
random value ${\bm V}$ on the internal time $\tau$. Recall that
$p^S(t,\tau)$ is the probability to be at the operational time
$\tau$ on the real time $t$. It is well known that the stochastic
differential equation of type (\ref{eq1}) is equivalent to the
corresponding FPE. In particular, the probability density
$p^V({\bm V}, \tau)$ obeys the standard FPE
\begin{displaymath}
\frac{\partial p^V({\bm V}, \tau)}{\partial\tau}={\hat L}_{\rm
FP}\, p^V({\bm V}, \tau),
\end{displaymath}
where ${\hat L}_{\rm FP}$ is a time-independent Fokker-Planck
operator whose exact form is not important here. The Laplace
transform of the function $p^{\,v_t}({\bm V}, t)$ with respect to
time replaces the integral relation (\ref{eq4}) by the algebraic
one, $\bar{p}^{\,v_t}({\bm V}, u)=u^{\alpha-1}\bar{p}^V({\bm V},
u^\alpha)$. Acting the operator ${\hat L}_{\rm FP}$ on the Laplace
image, we obtain $[{\hat L}_{\rm FP}\,\bar{p}^{\,v_t}]({\bm V},
u)= u^\alpha\,\bar{p}^{\,v_t}({\bm V}, u)-Q({\bm
V})\,u^{\alpha-1}$, where $Q({\bm V})$ is the initial condition.
The inverse Laplace transform gives the fractional FPE
\begin{eqnarray}
p^{\,v_t}({\bm V}, t)=&&Q({\bm V})+\nonumber\\
+\,\,\,\frac{1}{\Gamma(\alpha)}&&\!\!\!\int_0^td\tau\,
(t-\tau)^{\alpha-1}\,[{\hat L}_{\rm FP}\,p^{\,v_t}]({\bm V},
\tau).\label{eq5}
\end{eqnarray}
This equation has also the equivalent form
\begin{displaymath}
\frac{\partial^\alpha p^{\,v_t}({\bm V}, t)}{\partial t^\alpha}
-\frac{Q({\bm V})\,t^{-\alpha}}{\Gamma(1-\alpha)}={\hat L}_{\rm
FP}\,p^{\,v_t}({\bm V}, t)\,,
\end{displaymath}
where $\partial^\alpha/\partial t^\alpha$ denotes the
Liouville-Riemann fractional differential operator of order
$\alpha$ \cite{19}. Our analysis generalizes the mathematical
treatment of \cite{17} and shows that the Fokker-Planck operator
can have a more general form  rather than only with a temporally
constant force field. Another approach to the description of
anomalous transport in external fields is developed in \cite{20}.
The consideration is based on a generalization of the classical
Chapman-Kolmogorov equation. An interesting justification of the
generalized Chapman-Kolmogorov equation is that trapping events
are superimposed on the Langevin dynamics, with a waiting time
distribution with infinite mean. By the choice of special forms
for the transfer kernel and the probability density function of
the waiting time between any two successive jump events in the
generalized equation, one can recover some models discussed in the
literature.

If the probability density $p^V({\bm V}, \tau)$ is known
explicitly, the solution of (\ref{eq5}) can be calculated by mean
of
\begin{equation}
p^{\,v_t}({\bm V}, t)= \int^\infty_0F_\alpha(z)\, p^V({\bm V},
t^\alpha z)\,dz\,.\label{eq6}
\end{equation}
The formula is especially useful for some particular cases whose
the exact solutions of the ordinary FPE have a closed form (for
example, the harmonic potential leading to a linear force field in
FPE). It is interesting also to observe the integral
representation of the fractional FPE solution in \cite{41} (see
expression (2.32)). Now clearly, that formula is nothing else as a
consequence of the subordination relation (\ref{eq4}) (or
(\ref{eq6})). In this connection it should be noted that the
relation (\ref{eq4}) is not of convolution type, so the derivation
of Eq. (\ref{eq5}), having the fractional integral of time, from
the ordinary FPE is not entirely trivial.

Many papers \cite{28,35,37,38,39,40,41} have focused on the
derivation of the fractional FPE with different potentials and its
solution. Starting with \cite{42}, the continuous time random walk
approach is very popular for that goal. However, only recently it
has been shown that the solutions are density functions of a
stochastic process \cite{43}. We support the latter point of view:
the problem of anomalous diffusion should be analyzed with the
exact definition of the corresponding random process. The density
function and the master equation are derived from this process.

\section{Fluctuation-dissipation relation and
H-theorem}\label{par4}

According to Eq. (\ref{eq1}), the variance of the random variable
$\bm V$ is
\begin{equation}
\langle V^2_i(\tau)\rangle=v^2_{i,\,
0}\,e^{-2\gamma\tau}+\frac{D}{\gamma} \left(1-e^{-2\gamma\tau}
\right),\label{eq7}
\end{equation}
where $v_{i, 0}$ is the initial condition. Since the random
processes $\bm V$ and $S(t)$ are independent, we average the
expression (\ref{eq7}) on the internal variable $\tau$ so that
\begin{equation}
\overline{\langle V^2_i(t)\rangle}=\int^\infty_0F_\alpha(z)\,
\langle V^2_i(t^\alpha z)\rangle\,dz\,, \label{eq8}
\end{equation}
where the line over a variable denotes the average over the
internal variable $\tau$. Calculating the following integral
\begin{equation}
\int_0^\infty F_\alpha(z)\,e^{-2\gamma t^\alpha z}\,dz=
E_\alpha(-2\gamma t^\alpha)\,, \label{eq9}
\end{equation}
the exponential functions in (\ref{eq7}) are replaced with the
Mittag-Leffler function for (\ref{eq8}). The stationary state of
(\ref{eq3}) is finite so that
\begin{equation}
\lim_{t\to\infty}\overline{\langle V^2_i(t)\rangle}=D/\gamma\,.
\label{eq10}
\end{equation}
The constants $D$ and $\gamma$ are interpreted as generalized
diffusion and damping coefficients respectively. The mean
$\overline{\langle V_i(t)\rangle}$ is zero as well as $\langle
V_i(\tau)\rangle=0$. The boundary case $\alpha=1$ may be also
included in the study, as $S(t)=t$. The probability density
$p^S(t,\tau)$ reduces to the Dirac $\delta$-function, and Eq.
(\ref{eq1}) becomes the ordinary Langevin equation in the true
time $t$.

In the other hand, the energy of a classical system is distributed
equally among all degrees of freedom. We get the
fluctuation-dissipation relation $D/\gamma=k_BT/m$ for the given
temperature of a bath $T$ and the mass of a particle $m$, and
$k_B$ is the Boltzmann constant. The expression is very similar to
the Einstein relation, but the constants $D$ and $\gamma$ are
generalized. It should be pointed out that the stochastic time
arrow does not break this equal distribution law and influences
only on the character of relaxation (slow power decay). Therefore,
in this case the concept of temperature is valid, i.\ e.\ the
stationary state of the fractional FPE is defined by the
temperature $T$. The difference of entropies of equilibrium and
arbitrary states gives a Lyapunov functional $\Lambda(t)\geq 0$.
No wonder that its temporal evolution confirms the H-theorem.
Although the law of relaxation toward thermal equilibrium changes,
it remains monotonic, and the equilibrium state has the most
entropy due to the Gibbs-Boltzmann distribution. The fact, that no
modifications of the Boltzmann thermodynamics for anomalous
diffusion described by the equation of type (\ref{eq5}) are
required, was already noted in \cite{39,41,44}. However, the true
cause of the result was not established. Now it is clear that both
processes (\ref{eq1}) and (\ref{eq3}) are closely connected and
have a common ground.

\section{General kinetic equation with the stochastic time
clock}\label{par5} For a general type of a Markovian process the
general kinetic equation is
\begin{equation}
\frac{dp_n(t)}{dt}=\sum_{k=0}^{\infty}\left\{W_{nk}p_k(t)-
W_{kn}p_n(t)\right\}\,, \label{eq11}
\end{equation}
where $W_{kn}$ are the transition probability rates from state $n$
to state $k$. This equation defines the probability $p_n$ for the
system to be in state $n$. The term $W_{nk}p_k$ describes
transitions into the state $n$ from states $k$, and $W_{kn}p_n$
corresponds to transition out of $n$ into other states $k$. The
continuous version of Eq. (\ref{eq11}) takes the form
\begin{displaymath}
\frac{dP(y,t)}{dt}=\int\left\{W(y\mid y')P(y',t)-W(y'\mid
y)P(y,t)\right\}dy'\,
\end{displaymath}
with the initial condition $P(y,0)$. Let us represent both these
equations as
\begin{equation}
\frac{dp\,(t)}{dt}=\hat{\bm W}p\,(t)\,, \label{eq12}
\end{equation}
where $\hat{\bm W}$ denotes the transition rate operator. It is
important to emphasize here that this operator is
time-independent. The equation (\ref{eq12}) can be written in the
integral form
\begin{displaymath}
p\,(t)=p\,(0)+\int^t_0d\tau\,\hat{\bm W}p\,(\tau)\,.
\end{displaymath}
The Laplace transform $\tilde p\,(s)$ with respect to $t$ is given
by
\begin{displaymath}
\tilde p\,(s)=\int_0^\infty e^{-st}\,p\,(t)\,dt
\end{displaymath}
and leads to
\begin{displaymath}
\hat{\bm W}\,\tilde p\,(s)=s\tilde p\,(s)-p\,(0)\,.
\end{displaymath}
Now we determine a new process with the probability equal to
\begin{displaymath}
p_\alpha(t)=\int_0^\infty p^S(t,\tau)\,p\,(\tau)\,d\tau\,.
\end{displaymath}
In Laplace space the probabilities $p_\alpha(t)$ and $p\,(t)$ are
related by $\tilde p_\alpha(s)=s^{\alpha-1}\tilde p\,(s^\alpha)$,
where
\begin{displaymath}
\tilde p_\alpha(s)=\int_0^\infty e^{-st}\,p_\alpha(t)\,dt
\end{displaymath}
is the Laplace image of $p_\alpha(t)$. By the simple algebraic
transformations we find
\begin{eqnarray}
\hat{\bm W}\,\tilde p_\alpha(s)&=&s^{\alpha-1}\,\hat{\bm
W}\,\tilde p\,(s^\alpha)\nonumber\\
&=&s^{\alpha-1}\,\{s^\alpha\,\tilde
p\,(s^\alpha)-p\,(0)\}\nonumber\\  &=&s^\alpha\tilde
p_\alpha(s)-p\,(0)\,s^{\alpha-1}\,. \label{eq13}
\end{eqnarray}
Thus, the fractional extension of Eq. (\ref{eq12}) reads
\begin{equation}
p_\alpha(t)=p\,(0)+\frac{1}{\Gamma(\alpha)}\int^t_0d\tau\,(t-
\tau)^{\alpha-1}\hat{\bm W}\,p_\alpha(\tau)\,. \label{eq14}
\end{equation}
For $\alpha=1$ we recover Eq. (\ref{eq12}). For a system with
discrete states the generating function is of the form
\begin{displaymath}
G(\zeta, t)=\sum_{k=0}^\infty\zeta^kp_k(t)\,,
\end{displaymath}
where the restriction $\mid\zeta\mid\leq 1$ is imposed to ensure
convergence of the series. With the help of the generating
function, one can find the moments by taking the derivative with
respect to $\zeta$ and then setting $\zeta=1$. The generating
function  of the process governed by the stochastic time clock is
given by the relation
\begin{equation}
G_\alpha(\zeta, t)=\int^\infty_0F_\alpha(z)\,G(\zeta, t^\alpha
z)\,dz\,. \label{eq15}
\end{equation}
Thus, the generating function for a discrete Markov process
directed by the process $S(t)$ can be obtained from the
appropriate generating function of the parent process by immediate
integration.

\begin{figure}
\includegraphics[width=8.6 cm]{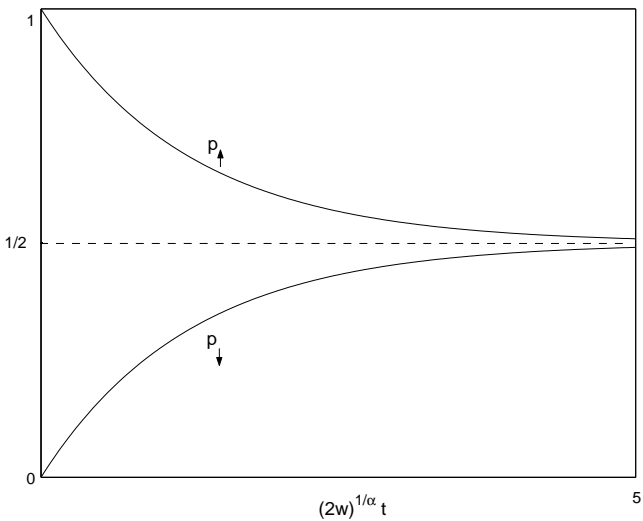}
\caption{\label{fig2}The relaxation of the probabilities
$p_\uparrow$ [Eq.(\ref{eq16})] and $p_\downarrow$
[Eq.(\ref{eq17})] plotted as a function of time. The dotted line
corresponds to the equilibrium state.}
\end{figure}

As an example, we consider the relaxation in a two-state system.
Let $N$ be the common number of objects in this system. If
$N_\uparrow$ is the number of objects in the state $\uparrow$,
$N_\downarrow$ is the number of objects in the state $\downarrow$
so that $N=N_\uparrow+N_\downarrow$. Assume that for $t=0$ the
states $\uparrow$ dominate, i.\ e.\
\begin{displaymath}
\frac{N_\uparrow(t=0)}{N}=p_\uparrow(0)=1,\quad
\frac{N_\downarrow(t=0)}{N}=p_\downarrow(0)=0\,,
\end{displaymath}
where $p_\uparrow$ and $p_\downarrow$ are the probabilities to
find the system in the states $\uparrow$ and $\downarrow$
respectively. Denote the transition rates by $w$. In the case the
general kinetic equation with the stochastic time clock
(\ref{eq14}) is written as
\begin{eqnarray}
p_\uparrow(t)&=&p_\uparrow(0)+\frac{w}{\Gamma(\alpha)}\int^t_0(t-
\tau)^{\alpha-1}\,\{p_\downarrow(\tau)-p_\uparrow(\tau)\}d\tau,\nonumber\\
p_\downarrow(t)&=&p_\downarrow(0)+\frac{w}{\Gamma(\alpha)}\int^t_0(t-
\tau)^{\alpha-1}\,\{p_\uparrow(\tau)-p_\downarrow(\tau)\}d\tau.\nonumber
\end{eqnarray}
From the linearity of these equations it follows
\begin{eqnarray}
p_\uparrow(t)+p_\downarrow(t)&=&p_\uparrow(0)+p_\downarrow(0)\,,
\quad p_\uparrow(t)-p_\downarrow(t)=\nonumber\\
p_\uparrow(0)-p_\downarrow(0)&-&\frac{2w}{\Gamma(\alpha)}
\int^t_0(t-\tau)^{\alpha-1}\,\{p_\uparrow(\tau)-
p_\downarrow(\tau)\}d\tau.\nonumber
\end{eqnarray}
Consequently, we obtain
\begin{eqnarray}
p_\uparrow(t)=\frac{1}{2}+\frac{1}{2}E_\alpha(-2wt^\alpha),\label{eq16}\\
p_\downarrow(t)=\frac{1}{2}-\frac{1}{2}E_\alpha(-2wt^\alpha)\label{eq17}.
\end{eqnarray}
The steady state of the system corresponds to equilibrium,
$p_\uparrow(\infty)=p_\downarrow(\infty)=1/2$ (Fig. \ref{fig2}).
The transition rate $w$ is defined by microscopic properties of
the system (for instance, from the given Hamiltonian of
interaction and the Fermi's golden rule). The value
$(2w)^{-1/\alpha}$ may be interpreted as a generalized relaxation
time. The randomization of time clock essentially changes the
character of relaxation in such a two-state system. If only
$\alpha\neq 1$, the relaxation has an algebraic decay. In this
connection it should be mentioned here that the experimental
relaxation curves of glasses show just the algebraic decay
\cite{45}.

\section{Summary}\label{par6}
We have shown that the fractional FPE can be derived by using the
ordinary Langevin equation. Although the processes described by
Eq. (\ref{eq5}) are non-Markovian at the true time, they are
Markovian with regard to the internal time. So the strange
kinetics results in the randomization of time clock of a Markov
process. In the probability theory the operation is called the
subordination of one random process by another \cite{15a}. As has
been stated above, the subordination does not break the
fluctuation-dissipation relation and the H-theorem. The stochastic
time clock has a clear physical sense -- a particle interacts with
a bath in random points of time so that there are memory effects.
It should be noted that the memory is a direct consequence of the
random time steps belonging to the strict domain of attraction of
a $\alpha$-stable distribution. One and only one index $\alpha$
characterizes both the corresponding $\alpha$-stable process and
its hitting time process. The stochastic differential equation
(\ref{eq3}) describes a random velocity field directed by a random
Markov process. In this case the dynamical foundation of
statistical physics is valid. The procedure of the randomization
of time clock extends the domain of applicability for the general
kinetic equation.

\acknowledgments The author thanks the referees for their useful
comments.

\bibliographystyle{apsrev}
\bibliography{stanislavsky}

\end{document}